# Aggregation in Probabilistic Databases via Knowledge Compilation

Robert Fink and Larisa Han and Dan Olteanu
Deptartment of Computer Science, University of Oxford
Wolfson Building, Parks Road, OX1 3QD Oxford, UK
{robert.fink,dan.olteanu}@cs.ox.ac.uk, hanlarisa@gmail.com

## ABSTRACT

This paper presents a query evaluation technique for positive relational algebra queries with aggregates on a representation system for probabilistic data based on the algebraic structures of semiring and semimodule. The core of our evaluation technique is a procedure that compiles semimodule and semiring expressions into so-called decomposition trees, for which the computation of the probability distribution can be done in time linear in the product of the sizes of the probability distributions represented by its nodes. We give syntactic characterisations of tractable queries with aggregates by exploiting the connection between query tractability and polynomial-time decomposition trees.

A prototype of the technique is incorporated in the probabilistic database engine SPROUT. We report on performance experiments with custom datasets and TPC-H data.

## 1. INTRODUCTION

This paper considers the evaluation problem for queries with aggregates on probabilistic databases.

The utility of aggregation has been argued for at length. In particular, aggregates are crucial for OLAP and decision support systems. All 22 TPC-H queries involve aggregation.

Probabilistic databases are useful to represent and query imprecise and uncertain data, such as data acquired through measurements, integrated from multiple sources, or produced by information extraction [21]. In this paper, we use a representation system for probabilistic data called *pvc-tables*. It is based on the algebraic structures of semiring and semimodule to support a mixed representation of aggregated values and tuple annotations for different classes of annotations and aggregations [2]. The pvc-tables can represent any finite probability distribution over relational databases. In addition, the results of queries with aggregates can be represented as pvc-tables of polynomial size. This contrasts with main-stream representation systems such as pc-tables [21], which can require an exponential-size overhead [15].

The problem of query evaluation is #P-hard already for simple conjunctive queries [21]. Aggregates are a further source of computational complexity: for example, already deciding whether there is a possible world in which the SUM of values of an attribute equals a given constant is NP-hard. Existing approaches to aggregates in probabilistic databases have considered restricted instances of the problem: they focus on aggregates over one probabilistic table of restricted expressiveness [4, 20, 16], or rely on expected values and Monte-Carlo sampling [10, 12, 22]. Expected values can lead to unintuitive query answers, for instance when data values and their probabilities follow skewed and non-aligned distributions [19]. Abiteboul et al. investigate XML queries with aggregates on probabilistic data [1]. An algebra proposed by Koch represents annotations and data values as rings which enables efficient incremental view maintenance in the presence of aggregations [13].

Our approach considers the problem of exact probability computation for positive relational algebra queries with aggregates on pvc-tables. The core of our technique is a procedure that compiles arbitrary semimodule and semiring expressions over random variables into so-called decomposition trees, for which the computation of the probability distribution can be done in polynomial time in the size of the tree and of the distributions at its nodes. Decomposition trees are a knowledge compilation technique [5] that reflects structural decompositions of expressions into independent and mutually exclusive sub-expressions. Flavours of decomposition trees have been proposed as compilation target for propositional formulas that arise in the evaluation of relational algebra queries (without aggregates) on probabilistic c-tables [18]. It has been shown that more complex tasks, such as conditioning probabilistic databases on given constraints [14] and sensitivity analysis and explanation of query results [11], can benefit from decomposition trees.

EXAMPLE 1. Figure 1 shows six pvc-tables, amongst them the suppliers table $S$, the products tables $P_1$ and $P_2$, and the table $PS$ pairing suppliers and products. They all have an annotation column $\Phi$ to hold expressions in a *semiring* $K$ generated by a set of independent random variables, with operations sum $(+)$ and product $(\cdot)$, and neutral elements $0_k$ and $1_K$. Each valuation of the random variables into a semiring (e.g. integers or Booleans) canonically maps semiring expressions into that semiring by interpreting $+$ and $\cdot$ as the corresponding operations of that semiring. Each such valuation defines a possible world of the database.

Figure 1d shows the result of the query $Q_1$ that asks for prices of products available in shops. The annotations of the result tuples are constructed as follows: The annotation of a join of two tuples is the product of their annotations, and the annotation obtained from projection or union is the sum of

| $S$ | | | $PS$ | | | | $P_1$ | | | $Q_1 = \pi_{\text{shop, price}}[S \bowtie PS \bowtie (P_1 \cup P_2)]$ | | | $Q_2 = \pi_{\text{shop}} \sigma_{P \leq 50} \varpi_{\text{shop};P \leftarrow \max(\text{price})}[Q_1]$ | |
|---|---|---|---|---|---|---|---|---|---|---|---|---|---|---|
| sid | shop | $\Phi$ | sid | pid | price | $\Phi$ | pid | weight | $\Phi$ | shop | price | $\Phi$ | Shop | $\Phi$ |
| 1 | M&S | $x_1$ | 1 | 1 | 10 | $y_{11}$ | 1 | 4 | $z_1$ | M&S | 10 | $x_1 y_{11}(z_1 + z_5)$ | M&S | $[x_1 y_{11}(z_1 + z_5) \otimes 10 +_{\max}$ |
| 2 | M&S | $x_2$ | 1 | 2 | 50 | $y_{12}$ | 2 | 8 | $z_2$ | M&S | 50 | $x_1 y_{12} z_2$ | | $x_1 y_{12} z_2 \otimes 50 +_{\max}$ |
| 3 | M&S | $x_3$ | 2 | 1 | 11 | $y_{21}$ | 3 | 7 | $z_3$ | M&S | 11 | $x_2 y_{21}(z_1 + z_5)$ | | $x_2 y_{21}(z_1 + z_5) \otimes 11 +_{\max}$ |
| 4 | Gap | $x_4$ | 2 | 2 | 60 | $y_{22}$ | 4 | 6 | $z_4$ | M&S | 60 | $x_2 y_{22} z_2$ | | $x_2 y_{22} z_2 \otimes 60 +_{\max}$ |
| 5 | Gap | $x_5$ | 3 | 3 | 15 | $y_{33}$ | | | | M&S | 15 | $x_3 y_{33} z_3$ | | $x_3 y_{33} z_3 \otimes 60 +_{\max}$ |
| | | | 3 | 4 | 40 | $y_{34}$ | | $P_2$ | | M&S | 40 | $x_3 y_{34} z_4$ | | $x_3 y_{34} z_4 \otimes 15 \leq 50] \cdot \Psi_1$ |
| | | | 4 | 1 | 15 | $y_{41}$ | pid | weight | E | Gap | 15 | $x_4 y_{41}(z_1 + z_5)$ | Gap | $[x_4 y_{41}(z_1 + z_5) \otimes 15 +_{\max}$ |
| | | | 4 | 3 | 60 | $y_{43}$ | 1 | 5 | $z_5$ | Gap | 60 | $x_4 y_{43} z_3$ | | $x_4 y_{43} z_3 \otimes 60 +_{\max}$ |
| | | | 5 | 1 | 10 | $y_{51}$ | | | | Gap | 10 | $x_5 y_{51}(z_1 + z_5)$ | | $x_5 y_{51}(z_1 + z_5) \otimes 10 \leq 50] \cdot \Psi_2$ |
| (a) | | | (b) | | | | (c) | | | (d) | | | (e) | |

$\Psi_1 = [x_1 y_{11}(z_1 + z_5) + x_1 y_{12} z_2 + x_2 y_{21}(z_1 + z_5) + x_2 y_{22} z_2 + x_3 y_{33} z_3 + x_3 y_{34} z_4 \neq 0_K]$  $\Psi_2 = [x_4 y_{41}(z_1 + z_5) + x_4 y_{43} z_3 + x_5 y_{51}(z_1 + z_5) \neq 0_K]$

**Figure 1: A database containing relations (a) Supplier $S$, (c) Products $P_1$, $P_2$, and (b) the many-to-many relation ProductSupplied $PS$, and the results of the positive query $Q_1$ and of the aggregate query $Q_2$.**

the annotations of the participating tuples [7]. For instance, the tuple $\langle \text{M\&S}, 10 \rangle$ has the annotation $x_1 y_{11}(z_1 + z_5)$, whose probability distribution can be computed as a function of probability distributions of the random variables $x_1, y_{11}, z_1$, and $z_5$ [21].

Consider the query $Q_2$ from Figure 1e that asks for shops in which the maximal price for the products in $P_1$ or $P_2$ is less than 50. Aggregation is expressed using the $\varpi$ operator, which in this query groups by the column shop and applies the aggregation MAX on price within each group.

The annotations of result tuples are built using *semimodule* expressions of the form $\Psi \otimes v$, where $\Psi$ is a semiring expression and $v$ is a data value. Such expressions can be "summed up" with respect to aggregation operations: For MIN, the sum $\alpha +_{\min} \beta$ is $\min(\alpha, \beta)$; for MAX, $\alpha +_{\max} \beta = \max(\alpha, \beta)$; for SUM, $\alpha +_{\text{sum}} \beta = \alpha + \beta$. The sums correspond to operations in *commutative monoids*. The annotation $\Phi$ of M&S in $Q_2$'s result is constructed as follows. This tuple represents a group of six tuples in the result of $Q_1$, all with the M&S shop value. The annotation $\Phi$ then expresses the conditions (1) that the sum of the price values of these six tuples in the MAX monoid is less than 50, and (2) that the group is not empty (as expressed by $\Psi_1$). Depending on the valuation of the variables in $\Phi$, these conditions can be true ($\top$) or false ($\bot$), or, more generally, the additive or multiplicative neutral element of the semiring.

For instance, a valuation $\nu_1$ that maps $x_1, x_2, y_{11}, y_{21}, z_1, z_2, z_5$ to $\top$ and all other variables to $\bot$ satisfies $\Phi$, since

$\nu_1(\Phi) \equiv [\top \otimes 10 +_{\max} \bot \otimes 50 +_{\max} \top \otimes 11 +_{\max}$
$\bot \otimes 60 +_{\max} \bot \otimes 60 +_{\max} \bot \otimes 15 \leq 50] \cdot \top$
$\equiv [10 +_{\max} 11 \leq 50] \equiv [\max(10,11) \leq 50] \equiv \top.$  □

If the variables in such expressions are random variables, then the expressions themselves can be interpreted as random variables. Moreover, the probability distributions of the obtained expressions reflect the probabilities of query answers taking particular values in a randomly drawn world of the database. Our technique allows to efficiently compute probabilities defined by such expressions by structural decomposition. For example, an expression $\alpha = ab \otimes 10 + xy \otimes 20$ can be decomposed in independent sub-expressions $ab \otimes 10$ and $xy \otimes 20$ that do not share variable symbols and hence constitute independent random variables.

The structure of the paper follows the list of contributions:

- We present an evaluation framework for queries with aggregates (SUM, PROD, COUNT, MIN, MAX) on pvc-tables, a representation system for probabilistic data based on semirings and semimodules.

- We devise a technique for computing the exact probability distribution of query results based on a generic compilation procedure of *arbitrary* semimodule and semiring expressions into so-called decomposition trees, for which the computation of the probability distribution can be done in time linear in the product of the sizes of the distributions represented by its nodes.

- We give a syntactic characterisation of a class of aggregate queries that are tractable on tuple-independent databases. Our query tractability result follows from the observation that the semiring and semimodule expressions in the result of our tractable queries admit polynomial size decomposition trees and polynomial size probability distributions at their nodes.

- A prototype of our technique is incorporated into the probabilistic database engine SPROUT.

- Extensive performance experiments using our own synthetic datasets and TPC-H data are discussed.

Besides exact computation, decomposition trees also allow for approximate probability computation [18]. Due to lack of space, we refer the reader to the MSc thesis of the second author [9]. The pvc-tables can be extended to cope with continuous probability distributions, similar to the extensions of pc-tables in the PIP system [12].

## 2. PRELIMINARIES

### 2.1 Induced Discrete Probability Space

Let $S$ be a countable set and $\mathbf{X}$ be a finite set of $S$-valued independent random variables. We denote by $P_x$ the discrete probability distribution of a variable $x \in \mathbf{X}$, and by $P_x[s]$ the probability that $x$ takes value $s \in S$; we often specify $P_x$ by the set of pairs of unique values with their non-zero probabilities, $\{(s, P_x[s]) \mid s \in S \text{ and } P_x[s] > 0\}$. The *size* of a probability distribution is the size of its set representation.

DEFINITION 1. *Let $\Omega = \{\nu : \mathbf{X} \to S\}$ be the set of mappings from $\mathbf{X}$ into $S$. A probability mass function*

$$\Pr(\nu) = \prod_{x \in \mathbf{X}} P_x[\nu(x)]$$

*for every sample $\nu \in \Omega$, and a probability measure*

$$\Pr(E) = \sum_{\nu \in E} \Pr(\nu) \qquad \text{for all } E \subseteq \Omega$$

*define a probability space $(\Omega, 2^\Omega, \Pr)$ that we call the probability space induced by $\mathbf{X}$.*

The probability distribution of the sum of two independent random variables is the convolution of their individual distributions [8]. For instance, given two random variables $x, y$ over positive integers, the probability that the sum of the random variables equals to 4 is the sum of the probabilities of $x$ being 0 and $y$ being 4, of $x$ being 1 and $y$ being 3, and so on. The applicability of convolution to determine the probability distribution of a function of independent random variables is not limited to sums of integers:

PROPOSITION 1. *Given sets $A, B, C$, an $A$-valued random variable $x$ with probability distribution $P_x$, a $B$-valued random variable $y$ with probability distribution $P_y$, $x$ independent of $y$, and an operation $\bullet : A \times B \to C$, the probability distribution $P_{x \bullet y}$ of the $C$-valued random variable $z = x \bullet y$ is the convolution of $P_x$ and $P_y$ with respect to $\bullet$:*

$$P_{x \bullet y}[c] = \sum_{\substack{(a,b) \in A \times B: \\ c = a \bullet b}} P_x[a] \, P_y[b] \qquad \text{for all } c \in C \quad (1)$$

EXAMPLE 2. The formula for the probability $P_{\Phi \vee \Psi}$ of the disjunction $\Phi \vee \Psi$ of two independent Boolean random variables is a special case of Eq. (1):

$$P_{\Phi \vee \Psi}[\top] = \sum_{\substack{(a,b) \in \mathbb{B} \times \mathbb{B}: \\ \top = a \vee b}} P_\Phi[a] \, P_\Psi[b]$$
$$= P_\Phi[\top] P_\Psi[\top] + P_\Phi[\bot] P_\Psi[\top] + P_\Phi[\top] P_\Psi[\bot]$$
$$= 1 - (1 - P_\Phi[\top])(1 - P_\Psi[\top]) \qquad \square$$

REMARK 1. The sum in Eq. (1) is invariant under the restriction to those $a$ and $b$ for which $P_x[a] > 0$ and $P_y[b] > 0$. Hence, even if the cardinality of $A \times B$ is infinite, the convolution is a finite sum whenever only finitely many elements in $A$ and $B$ have non-zero probability. $\square$

## 2.2 Monoids, Semirings and Semimodule

Our representation system for probabilistic data is based on the notions of monoid, semiring, and semimodule.

DEFINITION 2. *A* monoid *is a set $M$ with an operation $+ : M \times M \to M$ and a neutral element $0 \in M$ that satisfy the following axioms for all $m_1, m_2, m_3 \in M$:*

$$(m_1 + m_2) + m_3 = m_1 + (m_2 + m_3)$$
$$0 + m_1 = m_1 + 0 = m_1$$

*A monoid is* commutative *if $m_1 + m_2 = m_2 + m_1$.*

Monoids naturally describe many aggregation operations. Aggregation over a column of a relation fixes a domain of values, usually $\mathbb{R}$ or $\mathbb{N}$, and a binary operation. For instance, the MIN of a column with values $v_1, \cdots, v_n \in \mathbb{R}$ is

$$\min(v_1, \cdots, v_n) = \min(v_1, \min(v_2, \cdots \min(v_{n-1}, v_n) \cdots)).$$

The binary operation is *commutative* and *associative*, i.e., the value of the aggregation is invariant under the order in which it is computed. Furthermore, each aggregation operation has a *neutral element*, i.e., a value that does not contribute to the aggregation. For example, $0 \in \mathbb{R}$ is the neutral element for SUM, and $\infty$ is the neutral element for MIN. These aggregations correspond to commutative monoids, in particular: SUM = $(\mathbb{N}, +, 0)$, MIN = $(\mathbb{N}^{\pm \infty}, \min, +\infty)$, MAX = $(\mathbb{N}^{\pm \infty}, \max, -\infty)$. COUNT is a special case of SUM. More complicated aggregations (e.g., AVG) can conceptually be composed from simpler ones (e.g., SUM and COUNT), but their treatment is out of the scope of this paper.

DEFINITION 3. *A* commutative semiring *is a set $S$ together with operations $+, \cdot : S \times S \to S$ and neutral elements $0, 1 \in S$ such that $(S, +, 0)$ and $(S, \cdot, 1)$ are commutative monoids and the following holds for all $s_1, s_2, s_3 \in S$:*

$$s_2 \cdot (s_2 + s_3) = (s_1 \cdot s_2) + (s_1 \cdot s_3)$$
$$(s_1 + s_2) \cdot s_3 = (s_1 \cdot s_3) + (s_2 \cdot s_3)$$
$$0 \cdot s_1 = s_1 \cdot 0 = 0.$$

Commutative semirings are the canonical algebraic structure for tuple annotations [7]. Annotations from the Boolean semiring yield set semantics, annotations from $\mathbb{N}$ correspond to bag semantics, and annotations from the security semiring can be used to constrain access to query results depending on access rights to database tuples that contributed to the result [2]. The most general semirings are those generated over a set of variables. Intuitively, the carrier of such semirings are syntactic expressions built from the variables and the multiplication and sum symbols, where elements are identified via the semiring laws. For example, given a set $\mathbf{X} = \{x_1, x_2, x_3\}$ of variables, the elements of the semiring PosBool($\mathbf{X}$) are positive Boolean expressions, e.g. $x_1 + x_2$ or $x_1(x_2 + x_3)$. By the distributivity law in semirings, the expressions $x_1(x_2 + x_3)$ and $x_1 x_2 + x_1 x_3$ are equal. A more general freely generated semiring is the ring of polynomials (also called free commutative algebra) over a set $\mathbf{X}$ of variables. Elements of generated semirings are called *semiring expressions* which we denote by $K$ throughout this paper.

DEFINITION 4. *Let $(S, +_S, 0_S, \cdot_S, 1_S)$ be a commutative semiring. An $S$-semimodule $M$ consists of a commutative monoid $(M, +_M, 0_M)$ and a binary operation $\otimes : S \times M \to M$ such that for all $s_1, s_2 \in S$ and $m_1, m_2 \in M$ we have*

$$s_1 \otimes (m_1 +_M m_2) = s_1 \otimes m_1 +_M s_1 \otimes m_2$$
$$(s_1 +_S s_2) \otimes m_1 = s_1 \otimes m_1 +_M s_2 \otimes m_1$$
$$(s_1 \cdot_S s_2) \otimes m_1 = s_1 \otimes (s_2 \otimes m_1)$$
$$s_1 \otimes 0_M = 0_S \otimes m_1 = 0_M$$
$$1_S \otimes m_1 = m_1.$$

We write $S \otimes M$ to denote a $S$-semimodule $M$, and write $\cdot$ for $\cdot_S$ and $+$ for $+_S, +_M$ whenever the meaning is unambiguous. Semimodules combine monoids with semirings to represent aggregation values *conditioned* on the value of a semiring expression. Analogous to the case of semiring expressions that correspond to freely generated semirings, we denote by *semimodule expressions* the elements of the $K$-semimodule generated by a given monoid. The semimodules we use frequently are $\mathbb{N} \otimes \mathbb{N}$ and $\mathbb{B} \otimes \mathbb{N}$ for MIN, MAX, SUM, and PROD monoids. A semimodule $\mathbb{B} \otimes \mathbb{N}$ over SUM would not have the intuitive semantics; this reflects the well-known incompatibility of SUM aggregation with set semantics.

## 2.3 Query Language

The query language under consideration is a restriction of positive relational algebra extended by an operator $\varpi$ for aggregation and grouping. Given a relation $R$ over schema $\Sigma$, the operator $\varpi$ in the query

$$\varpi_{\bar{A}; \alpha_1 \leftarrow \text{AGG}_1(B_1), \ldots, \alpha_l \leftarrow \text{AGG}_l(B_l)}(R),$$

where $(\bar{A} \cup \{B_1, \ldots, B_l\}) \subseteq \Sigma$, groups by the attributes $\bar{A}$ and takes the aggregations $\text{AGG}_1$ to $\text{AGG}_l$ over the attributes $B_1$ to $B_l$ respectively. The result has schema $\bar{A} \cup \{\alpha_1, \ldots, \alpha_l\}$, where $\alpha_1$ to $\alpha_l$ are *aggregation attributes*. We

$$\begin{aligned}
\Phi &::= x \mid \Phi + \Phi \mid \Phi \cdot \Phi \mid [\alpha\,\theta\,\alpha] \mid [\Phi\,\theta\,\Phi] \mid s \\
\alpha &::= \Phi \otimes m \; \{+_{op} \Phi \otimes m\} \mid m \\
op &::= \min \mid \max \mid \text{count} \mid \text{sum} \mid \text{prod} \\
\theta &::= = \mid \neq \mid \leq \mid \geq \mid < \mid > \\
x &::= \text{A variable symbol } x \in \mathbf{X} \\
m &::= \text{A value from an aggregation monoid } M \\
s &::= \text{A value from the semiring } S
\end{aligned}$$

Figure 2: Grammar for semiring expressions $\Phi \in K$ and semimodule expressions $\alpha \in (K \cup S) \otimes M$.

consider SUM, PROD, COUNT, MAX, and MIN aggregations, and restrict the queries such that projection, union and grouping are never applied to aggregation attributes. This restriction simplifies query rewriting, but is not requisite for our query evaluation method. Out of the 22 TPC-H queries, only query Q13 violates the restriction.

DEFINITION 5. *The query language $\mathcal{Q}$ considered in this paper consists of queries that are built using the relational operators $\delta, \sigma, \pi, \times, \cup, \varpi$ and satisfy the following constraints:*

1. *In $\pi_{\bar{A}}(Q)$ and $\varpi_{\bar{A};\alpha_1 \leftarrow AGG_1(B_1),\ldots,\alpha_l \leftarrow AGG_l(B_l)}(Q)$, the attributes in $\bar{A}$ are not aggregation attributes.*

2. *In $Q_1 \cup Q_2$, the attributes of $Q_1$ and $Q_2$ are not aggregation attributes.*

EXAMPLE 3. The TPC-H query Q1 has the structure `SELECT A,SUM(B) FROM R GROUP BY A` which is equivalent to $\varpi_{A;\beta \leftarrow \text{SUM}(B)}(R)$. TPC-H query Q2 has the structure `SELECT A FROM R WHERE B = (SELECT MIN(C) FROM S)`, or, equivalently, $\pi_A \sigma_{B=\gamma}(R \times \varpi_{\emptyset;\gamma \leftarrow \text{MIN}(C)}(S))$.

The query $R \cup \varpi_{A;\beta \leftarrow \text{SUM}(B)}(S)$ is not in $\mathcal{Q}$, since the second union term is a relation with the aggregation attribute $\beta$ and hence violates constraint 2 in Def. 5. However, $\pi_A(R) \cup \pi_A \sigma_{\beta \geq 5}(\varpi_{A;\beta \leftarrow \text{SUM}(B)}(S))$ is a valid $\mathcal{Q}$-query. □

## 3. PVC-TABLES: A REPRESENTATION SYSTEM FOR PROBABILISTIC DATA

In this section we introduce a succinct and complete representation system for probabilistic data, which we call *probabilistic value-conditioned tables* or pvc-tables for short. This system is based on work in databases with provenance information [2, 7, 3]. The reason for using pvc-tables, as opposed to main-stream representation systems such as pc-tables (and special cases such as tuple-independent or BID tables) [21], is that pvc-tables fit naturally with aggregate queries: answers to aggregate queries on pvc-tables or even on pc-tables are representable as pvc-tables of size polynomial in the size of the input tables, while they may require an exponential overhead when represented as pc-tables [15]. Aggregation on pvc-tables is handled using a mixed representation of tuple annotations and aggregated values using the algebraic structures of semirings and semimodules.

DEFINITION 6. *A pvc-table $T$ over a probability space $\Omega$ induced by a set $\mathbf{X}$ of variables is a relation with an annotation column $\Phi$ holding semiring expressions over $\mathbf{X}$, and where the tuple values can be constants or semimodule expressions over $\mathbf{X}$. A pvc-database $D = \{T_1, \ldots, T_n\}$ is a set of pvc-tables over the same probability space $\Omega$.*

| $S_\mathbb{B}$ | | | $S_{\mathbb{N},1}$ | | | $S_{\mathbb{N},2}$ | | |
|---|---|---|---|---|---|---|---|---|
| sid | shop | $\Phi$ | sid | shop | $\Phi$ | sid | shop | $\Phi$ |
| 1 | M&S | $\bot$ | 1 | M&S | 2 | 1 | M&S | 0 |
| 2 | M&S | $\top$ | 2 | M&S | 1 | 2 | M&S | 1 |
| 3 | M&S | $\bot$ | 3 | M&S | 1 | 3 | M&S | 3 |
| 4 | Gap | $\bot$ | 4 | Gap | 7 | 4 | Gap | 7 |
| 5 | Gap | $\top$ | 5 | Gap | 2 | 5 | Gap | 2 |
| (a) | | | (b) | | | | | |

Figure 3: Three possible worlds of pvc-table $S$ in Figure 1 under two different semirings: $S_\mathbb{B}$ is annotated with expressions from the Boolean semiring, and $S_{\mathbb{N},1}$ and $S_{\mathbb{N},2}$ with integers.

*The semantics of $D$ is given by its possible worlds:*

$$\Big\{ \{\{\nu(t) \mid t \in T_1\}, \ldots, \{\nu(t) \mid t \in T_n\}\} \;\Big|\; \nu \in \Omega \Big\}.$$

*where the mapping $\nu$ is applied to all expressions in tuples $t$ and is identity for constants. Each mapping $\nu \in \Omega$ defines a possible world.*

Columns that hold semimodule expressions are called *aggregation columns*. Semimodule expressions over different monoids can co-exist in a pvc-table. The annotation column $\Phi$ hosts semiring expressions that are generated by the variable set $\mathbf{X}$, and conditional expressions representing comparisons of expressions and constants. The other columns of a pvc-table host constants and semimodule expressions. A grammar for such expressions is given in Figure 2. All these types of expressions can occur in pvc-tables representing the result of aggregate queries; we show how to compute these expressions for any $\mathcal{Q}$-query in Section 4.

EXAMPLE 4. Figure 1 shows six pvc-tables. The pvc-table $S$ has annotations from a semiring over the set $\mathbf{X}$ of variables $x_1, \ldots, x_5$. Figure 3 shows a few possible worlds for this pvc-table for different semirings. First, consider valuations of the variables $\mathbf{X}$ into the Boolean semiring $\mathbb{B}$. Possible world $S_\mathbb{B}$ in Figure 3a has probability $P_{x_1}[\bot] \cdot P_{x_2}[\top] \cdot P_{x_3}[\bot] \cdot P_{x_4}[\bot] \cdot P_{x_5}[\top]$. Since $\mathbb{B}$ has only the two elements $\bot, \top$, the number of possible worlds is $2^5$. Secondly, consider valuations of $\mathbf{X}$ into $\mathbb{N}$. Figure 3b shows two possible worlds, with respective probabilities $P_{x_1}[2] \cdot P_{x_2}[1] \cdot P_{x_3}[1] \cdot P_{x_4}[7] \cdot P_{x_5}[2]$ and $P_{x_1}[0] \cdot P_{x_2}[1] \cdot P_{x_3}[3] \cdot P_{x_4}[7] \cdot P_{x_5}[2]$. □

The pvc-tables $PS$, $P_1$, $P_2$, and $Q_1$ in Figure 1 contain no semimodule expressions and are in fact pc-tables [21], a simpler representation system where tuple values are constants and annotations are expressions from semirings generated by a set of random variables. In pvc-tables, however, semimodule expressions can appear as tuple values.

EXAMPLE 5. Consider an aggregation AGG on the weight column of relation $P_1$ in Figure 1. The semimodule expression representing the aggregated value is $\alpha = z_1 \otimes 4 +_{\text{AGG}} z_2 \otimes 8 +_{\text{AGG}} z_3 \otimes 7 +_{\text{AGG}} z_4 \otimes 6$, where the $+_{\text{AGG}}$ operator depends on the particular aggregation monoid AGG. □

We next look at the semantics of expressions under valuations of variables; this will be extended to a probabilistic interpretation in Section 5.

**Semiring, Monoid, and Semimodule Homomorphism.** Let $K$ be the semiring generated by $\mathbf{X}$; the variables in $\mathbf{X}$ are themselves elements of $K$, i.e. $\mathbf{X} \subseteq K$. Given another semiring $S$, a mapping $\nu : \mathbf{X} \to S$ of the variables can be uniquely extended to a semiring homomorphism $\nu : K \to S$ that "evaluates" semiring expressions from $K$ to elements in $S$.

| Database Semantics | | $S$ | Probability Distributions |
|---|---|---|---|
| Deterministic | Set | $\mathbb{B}$ | $P_x[\top] = 1$ or $P_x[\bot] = 1$ |
| Deterministic | Bag | $\mathbb{N}$ | $\exists n \in \mathbb{N} : P_x[n] = 1$ |
| Probabilistic | Set | $\mathbb{B}$ | $P_x[\top], P_x[\bot] \in [0,1]$ |
| Probabilistic | Bag | $\mathbb{N}$ | $\forall n \in \mathbb{N} : P_x[n] \in [0,1]$ |

**Table 1: Database semantics for different semirings $S$ and probability distributions.**

A similar construction holds for semimodule: For semimodule expressions $W$ and a monoid M, a mapping $\nu : \mathbf{X} \to S$ is extended to a monoid homomorphism $\nu : W \to M$.

EXAMPLE 6. Consider the semimodule expression

$$\alpha = xy \otimes 5 +_{\min} (x+z) \otimes 10$$

over the semiring generated by $\mathbf{X} = \{x,y,z\}$ and the monoid $(\mathbb{N}, +_{\min}, \infty)$. The map $\nu : \mathbf{X} \to \mathbb{N}$ evaluates variable symbols to the semiring of positive integers and induces a monoid homomorphism that maps $\alpha$ to positive integers. For instance, the mapping $\nu : x \mapsto 2, y \mapsto 3, z \mapsto 0$ yields

$$\nu(\alpha) = \nu(x)\nu(y) \otimes 5 +_{\min} (\nu(x)+\nu(z)) \otimes 10$$
$$= 6 \otimes 5 +_{\min} 2 \otimes 10 = 1 \otimes 5 +_{\min} \ldots +_{\min} 1 \otimes 5+_{\min}$$
$$1 \otimes 10 +_{\min} \ldots +_{\min} 1 \otimes 10 = 5 +_{\min} 10 = 5.$$

Similarly, the value of $\alpha$ in Example 5 depends on the particular target semiring $S$ and the valuation of the variables. For instance, $\alpha \mapsto 24$ for SUM aggregation and semiring $\mathbb{N}$ with $z_1, z_2 \mapsto 2$ and $z_3, z_4 \mapsto 0$. Also, $\alpha \mapsto 6$ for MIN aggregation and the Boolean semiring with $z_1 \mapsto \bot$ and $z_2, z_3, z_4 \mapsto \top$. If all variables are mapped to $0_S$, the query answer is $0_M$, i.e. 0 in case of SUM and $+\infty$ for MIN. □

**Conditional Expressions** have the form $[\alpha\theta\beta]$, where $\alpha$ and $\beta$ are semimodule expressions for (possibly different) aggregation monoids and semirings generated by $\mathbf{X}$, and $\theta$ is a binary relation. Such conditional expressions can appear as tuple annotations in pvc-tables. Given a valuation $\nu : \mathbf{X} \to S$ and (semiring or semimodule) expressions $\Phi, \Psi$, the semantics of $[\Phi\theta\Psi]$ is defined by extending $\nu$ via

$$\nu([\Phi\theta\Psi]) = \begin{cases} 1_S, & \text{if } \nu(\Phi) \ \theta \ \nu(\Psi) \\ 0_S, & \text{otherwise} \end{cases} \quad (2)$$

Comparisons operators $\leq, \geq$ for $\theta$ only make sense for ordered carriers. In the light of Eq. (2), we can equivalently see $[\Phi\theta\Psi]$ as a binary operation $S \times S \to S$ or $M \times M \to S$.

EXAMPLE 7. The annotations in the pvc-table $Q_2$ in Figure 1 contain conditional expressions representing comparisons between semimodule expressions and a constant from $M$ and semiring expressions and the constant $0_K$. □

**Set vs. Bag Semantics.** The pvc-tables system is generic enough to encode both deterministic and probabilistic databases with set and bag semantics. Table 1 summarises how different choices for the semiring $S$ and probability distributions for variables $x \in \mathbf{X}$ give rise to those semantics.

Under the Boolean semiring, annotation expressions are evaluated to $\bot$ or $\top$, denoting tuples *in* or *not in* the database instance, respectively. If in addition every variable $x \in \mathbf{X}$ has either $P_x[\bot] = 1$ or $P_x[\top] = 1$, then we have the case of deterministic databases with set semantics, i.e. there is exactly one possible world with non-zero probability.

For deterministic bag semantics, tuple annotations are evaluated to $\mathbb{N}$ and represent tuple multiplicities and for each variable $x \in \mathbf{X}$ there is exactly one $n \in \mathbb{N}$ for which $P_x[n] = 1$. For probabilistic databases with set semantics, every answer tuple has a probability for being true or false, while in the case of bag semantics, the pvc-table encodes a probability distribution over the multiplicity of its tuples.

Existing representation systems for probabilistic data are over the Boolean semiring, i.e., their semantics is set-based. While this choice is justified in the absence of aggregation (or for MIN/MAX), it is *not* for COUNT/SUM/PROD aggregates, which require bag semantics. In the latter case, we need a semiring for which there is a homomorphism into $\mathbb{N}$ [2]. The set-based probabilistic database model is subsumed by the probabilistic bag-based model: Every $x \in \mathbf{X}$ is an $\mathbb{N}$-valued random variable, yet the probability distributions $P_x$ are non-zero only for $0, 1 \in \mathbb{N}$.

## 4. QUERY EVALUATION I: COMPUTING THE TUPLES IN THE QUERY RESULT

Our approach to query evaluation of $\mathcal{Q}$-queries on pvc-tables has two logical steps: In the first step we compute the tuples in the query result, and in the second step we compute their probability distributions. We discuss the first step in this section and the second step in the next section.

The tuples in the result of a query $Q$ can have semimodule expressions as values and semiring expressions as annotations. The construction of such expressions can be done by a query that can be statically inferred from $Q$ [2]. Figure 4 gives a translation $[\![\cdot]\!]$ of $\mathcal{Q}$ queries into SQL queries that compute the tuples in the query results (we chose SQL here, but the translation target can also be $\mathcal{Q}$). The rewriting accounts for joint and alternative use of data: Joint use of data (as in join) corresponds to multiplication in the annotation semiring and alternative use of data (as in union or projection) corresponds to summation in the annotation semiring. Semimodule expressions are constructed for the aggregation and grouping operator $\varpi$. The translation of the operators rename, selection, projection, product, and union is the same as for pc-tables [21]. In case of $\varpi$, the translation depends on the presence of group-by attributes. The translation uses the custom SQL operators $\sum_{\text{AGG}}, \sum_K, \cdot_K$ and $\otimes$ to construct annotation expressions.

EXAMPLE 8. The query $\varpi_{\emptyset; \alpha \leftarrow \text{AGG}(\text{weight})}(P_1)$ from Example 5 is rewritten into `select` $\sum_{\text{AGG}}(\texttt{R}.\Phi \otimes \texttt{R.weight})$ `as` $\alpha$, $1_K$ `as` $\Phi$ `from ( select * from` $P_1$`) R`. The answer to the rewritten query is one tuple of value $\alpha = z_1 \otimes 4 +_{\text{AGG}} z_2 \otimes 8 +_{\text{AGG}} z_3 \otimes 7 +_{\text{AGG}} z_4 \otimes 6$. The annotation of this tuple is the constant $1_K$, stating that the tuple $\langle \alpha \rangle$ is the query answer in all possible worlds. The evaluation of $\alpha$ may however yield different outcomes in different worlds.

The query $\pi_\emptyset \sigma_{5 \leq \alpha}(\varpi_{\emptyset; \alpha \leftarrow \text{MIN}(\text{weight})}(P_1))$ is rewritten to `select` $\texttt{S}.\Phi \cdot [5 \leq \texttt{S}.\alpha]$ `as` $\Phi$ `from (select` $\sum_{\text{MIN}}(\texttt{R}.\Phi \otimes \texttt{R.weight})$ `as` $\alpha$, $1_K$ `as` $\Phi$ `from (select * from` $P_1$`) R) S`. This Boolean query asks for the probability of the minimum weight of articles being larger than 5. The result of this query is a single empty tuple with annotation $\Phi = 1_K \cdot [z_1 \otimes 4 +_{\min} z_2 \otimes 8 +_{\min} z_3 \otimes 7 +_{\min} z_4 \otimes 6 \leq 5]$. □

As seen in the above example, for aggregation without grouping the resulting semimodule value is annotated with $1_K$, i.e., it "exists" in every possible world. In case of grouping, the resulting tuples are additionally annotated with a conditional expression to enforce that the group is not empty and thus at least one tuple in the group has a annotation

$$[\![R]\!] = \text{select R.*, R.}\Phi \text{ from R}$$

$$[\![\delta_{B\leftarrow A}(Q)]\!] = \text{select R.*, R.A as B, R.}\Phi \text{ as }\Phi \text{ from } ([\![Q]\!]) \text{ R}$$

$$[\![\sigma_{A\theta B}(Q)]\!] = \text{select R.*, R.}\Phi \cdot_K [A\theta B] \text{ as }\Phi \text{ from } ([\![Q]\!]) \text{ R}$$

$$[\![\pi_{A_1,\ldots,A_n}(Q)]\!] = \text{select R.}A_1, \ldots, \text{R.}A_n, \sum\nolimits_K (\text{R.}\Phi) \text{ as }\Phi \text{ from } ([\![Q]\!]) \text{ R group by R.}A_1, \ldots, \text{R.}A_n$$

$$[\![Q_1 \times Q_2]\!] = \text{select R.*, S.*, R.}\Phi \cdot_K \text{S.}\Phi \text{ as }\Phi \text{ from } ([\![Q_1]\!]) \text{ R}, ([\![Q_2]\!]) \text{ S}$$

$$[\![Q_1 \cup Q_2]\!] = \text{select R.*, } \sum\nolimits_K (\text{R.}\Phi) \text{ as }\Phi \text{ from}\bigl(\text{select * from } ([\![Q_1]\!]) \text{ union all select * from } ([\![Q_2]\!])\bigr) \text{ R group by R.*}$$

$$[\![\varpi_{A_1,\ldots,A_n;\alpha_1\leftarrow\text{AGG}_1(B_1),\ldots,\alpha_l\leftarrow\text{AGG}_l(B_l)}(Q)]\!] = \text{select R.}A_1, \ldots, \text{R.}A_n, \Gamma_1 \text{ as }\alpha_1, \ldots, \Gamma_l \text{ as }\alpha_l,$$
$$\bigl[\bigl(\sum\nolimits_K \text{R.}\Phi\bigr) \neq 0_K\bigr] \text{ as }\Phi \text{ from } ([\![Q]\!]) \text{ R group by R.}A_1, \ldots, \text{R.}A_n$$

$$[\![\varpi_{\emptyset;\alpha_1\leftarrow\text{AGG}_1(B_1),\ldots,\alpha_l\leftarrow\text{AGG}_l(B_l)}(Q)]\!] = \text{select } \Gamma_1 \text{ as }\alpha_1, \ldots, \Gamma_l \text{ as }\alpha_l, 1_K \text{ as }\Phi \text{ from } ([\![Q]\!]) \text{ R}$$

$$\text{where } \Gamma_i = \begin{cases} \sum_{\text{AGG}_i}(\text{R.}\Phi \otimes \text{R.}B_i) & \text{if AGG}_i = \text{MIN,MAX,SUM,PROD} \\ \sum_{\text{SUM}}(\text{R.}\Phi \otimes 1) & \text{if AGG}_i = \text{COUNT} \end{cases}$$

**Figure 4:** Recursive algorithm $[\![\cdot]\!]$ for rewriting a $\mathcal{Q}$ query to account for computation of semiring ($K$) and semimodule expressions. We assume that R.*, S.* do not select column $\Phi$.

different from $0_K$ in a possible world. This has been exemplified in the introduction for the query $Q_2$. However, this conditional expression is not always necessary:

EXAMPLE 9. Consider the query
$$Q_2' = \pi_{\text{shop}}\sigma_{P\leq 50}\varpi_{\text{shop};P\leftarrow \text{MIN(price)}}[Q_1]$$
similar to $Q_2$ on the database in Figure 1. Under the Boolean semiring $K$, the variables can take values $\bot = 0_K$ and $\top = 1_K$. In a possible world with $x_1, x_2, x_3 \mapsto \bot$, $\langle$M&S$\rangle$ is not an answer since there is no supplier for the shop M&S in the input instance $S$. Its annotation evaluates to $[\infty \leq 50] \cdot \bot = \bot \cdot \bot = \bot$. Here, the conditional expression $\Psi_1$ from Figure 1 would not be necessary, since it is implied by the first conditional expression in the expression $\Phi$. In case of other monoids, such as MAX (as discussed in the introduction), SUM, or COUNT, $\Psi_1$ is necessary. □

The constraints imposed on the query language $\mathcal{Q}$ by Definition 5 simplify the rewriting $[\![\cdot]\!]$: Since projections and unions are only done on non-aggregate columns, the rewriting rules for those operators can assume the input tuples to be free of semimodule expressions.

**Closure of pvc-tables under $\mathcal{Q}$ queries.** Since pvc-tables generalise pc-tables and the latter are complete in the sense that any finite probability distribution over relational databases can be represented using pc-tables [21], it follows that pvc-tables also form a complete representation system. In particular, they are closed under $\mathcal{Q}$ queries, in the sense that for any input pvc-database, the result of any $\mathcal{Q}$ query can be represented as a pvc-table. The pvc-tables representing query results are also succinct (for a fixed $\mathcal{Q}$ query), since the translation $[\![\cdot]\!]$ only constructs SQL queries of size linear in the size of the input $\mathcal{Q}$ queries and the evaluation of such SQL queries is in polynomial time data complexity.

THEOREM 1 (COMPLETENESS AND SUCCINCTNESS).

1. *Any finite probability distribution over relational database instances can be represented by pvc-databases.*

2. *Given a pvc-database $\mathcal{D}$ and a fixed $\mathcal{Q}$ query $Q$, the query result $[\![Q]\!](\mathcal{D})$ can be represented as a pvc-table with size polynomial in the size of $\mathcal{D}$.*

In particular, pvc-tables can be exponentially more succinct than pc-tables. The key difference is that pvc-tables allow for values and annotations to be intertwined in semimodule expressions, which can encode exponentially many possible outcomes of an aggregation in polynomial space. In pc-tables, we would need to enumerate all these outcomes [15].

## 5. QUERY EVALUATION II: PROBABILITY COMPUTATION BY DECOMPOSITION

This section presents a novel approach to computing probability distributions for query results on pvc-tables which is equivalent to computing probability distributions of the results' semiring and semimodule expressions. Our approach is based on compiling expressions into a tractable form called decomposition trees, for which distributions can be computed efficiently. A key property of our approach is its generality, as it applies to different semirings and semimodules.

**Expressions as Random Variables.** We first show how semiring and semimodule expressions — including conditional expressions — in pvc-tables give rise to random variables. Let $S$ be a countable semiring and $M$ a monoid, $\mathbf{X}$ a set of $S$-valued random variables, and $\Omega$ the probability space induced by $\mathbf{X}$; $K$ and $K \otimes M$ are the sets of semiring and semimodule expressions over $\mathbf{X}$. A semiring expression $\Phi \in K$ can be seen as a random variable $\Phi : \Omega \to S$ by letting $\Phi : \nu \mapsto \nu(\Phi)$ and with probability distribution

$$P_\Phi[s] = \Pr(\{\nu \in \Omega \mid \nu(\Phi) = s\}) = \sum_{\substack{\nu \in \Omega: \\ \nu(\Phi)=s}} \Pr(\nu). \quad (3)$$

A semimodule expression $\alpha \in K \otimes M$ is a random variable $\alpha : \Omega \to M$ with $\alpha : \nu \mapsto \nu(\alpha)$ and probability distribution $P_\alpha$ defined similar to Eq. (3) for all $m \in M$. If $M$ is over $\mathbb{R}$, then $\alpha$ is an $\mathbb{R}$-valued random variable. However, since the random variables $\mathbf{X}$ and hence the probability space $\Omega$ are countable, the set of values that $\alpha$ may take is countable.

**Probability Distributions of Independent Expressions.** Two semiring or semimodule expressions are *(syntactically) independent* if their sets of variables are disjoint; independent expressions are independent random variables.

EXAMPLE 10. Let $\mathbf{X} = \{x, y, a, b, c\}$, $M = \mathbb{N}$, $\Phi = x + y$ and $\alpha = a(b+c) \otimes 10 + c \otimes 20$. Then $\Phi$ and $\alpha$ are independent random variables since their sets of variables are disjoint. □

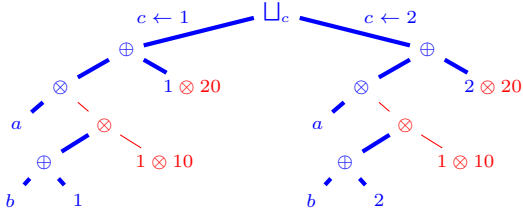

**Figure 5:** A d-tree for $\alpha = a(b+c) \otimes 10 + c \otimes 20$ over the semimodule $\mathbb{N} \otimes \mathbb{N}$. The node $\bigsqcup_c$ has one child for every $k \in \mathbb{N}$ with non-zero probability for $c \leftarrow k$, i.e. two children for $c = 1, 2$ in the setting of Ex.12. The thick (blue) part is a d-tree for the semiring component $a(b + c) + c$, where $\otimes$ is replaced by $\odot$.

The probability distribution of the sum or product of independent expressions is given by their convolution (cf. Proposition 1) with respect to the addition and multiplication operation of the semiring and semimodule. Given independent expressions $\Phi, \Psi \in K$ and $\alpha, \beta \in K \otimes M$ with probability distributions $P_\Phi, P_\Psi, P_\alpha, P_\beta$, we have for all $s \in S, m \in M$:

$$P_{\Phi+\Psi}[s] = \sum_{s_1, s_2 \in S : s_1 + s_2 = s} P_\Phi[s_1] \cdot P_\Psi[s_2] \quad (4)$$

$$P_{\Phi \cdot \Psi}[s] = \sum_{s_1, s_2 \in S : s_1 \cdot s_2 = s} P_\Phi[s_1] \cdot P_\Psi[s_2] \quad (5)$$

$$P_{\alpha+\beta}[m] = \sum_{m_1, m_2 \in M : m_1 + m_2 = m} P_\alpha[m_1] \cdot P_\beta[m_2] \quad (6)$$

$$P_{\Phi \otimes \alpha}[m] = \sum_{s_1 \in S, m_1 \in M : s_1 \otimes m_1 = m} P_\Phi[s_1] \cdot P_\alpha[m_1] \quad (7)$$

$$P_{[\alpha \theta \beta]}[s] = \sum_{m_1 \in M, m_2 \in M : [m_1 \theta m_2] = s} P_\alpha[m_1] \cdot P_\beta[m_2] \quad (8)$$

$$P_{[\Phi \theta \Psi]}[s] = \sum_{s_1 \in S, s_2 \in S : [s_1 \theta s_2] = s} P_\Phi[s_1] \cdot P_\Psi[s_2] \quad (9)$$

Following Remark 1, the sums in the above equations are restricted to summands of non-zero probabilities.

EXAMPLE 11. Let $S$ and $M$ be the semiring and the monoid of natural numbers with standard addition and multiplication. Let $\Phi = x$ with $P_x = \{(0, 0.3), (1, 0.3), (2, 0.4)\}$, and $\alpha = y \otimes 5$ be a semimodule expression with $P_y = \{(1, 0.4), (2, 0.4), (3, 0.2)\}$. Then $\alpha$ is a $M$-valued random variable with $P_\alpha = \{(5, 0.4), (10, 0.4), (15, 0.2)\}$. The probability distribution of $\Phi \otimes \alpha = x \otimes (y \otimes 5) = (x \cdot_S y) \otimes 5$ is given by Eq.(7). For example, $P_{\Phi \otimes \alpha}[10] = P_x[1]P_\alpha[10] + P_x[2]P_\alpha[5]$. The convolution is finite as the probability distributions are non-zero for finitely many elements. Further possible outcomes for $\Phi \otimes \alpha$ are 0, 5, 15, 20, 30. In case of the Boolean semiring, $S=\mathbb{B}$, possible outcomes are 0 and 5 with $P_{\Phi \otimes \alpha}[5] = P_x[\top]P_y[\top]$ and $P_{\Phi \otimes \alpha}[0] = 1 - P_{\Phi \otimes \alpha}[5]$. □

**Partitioning into Mutually Exclusive Expressions.** Given an expression $\Phi$ and a variable $x \in \mathbf{X}$ that occurs in $\Phi$, the probability distribution of $\Phi$ can be expressed using probability distributions of sub-expressions under valuations of $x$. For any $s' \in S$, we denote by $\Phi|_{x \leftarrow s'}$ the expression obtained from $\Phi$ by replacing every occurrence of $x$ by $s'$. Then the probability distribution of $\Phi$ can be partitioned by the probability distributions of expressions $\Phi|_{x \leftarrow s'}$:

$$P_\Phi[s] = \sum_{s' \in S} P_x[s'] \cdot P_{\Phi|_{x \leftarrow s'}}[s]. \quad (10)$$

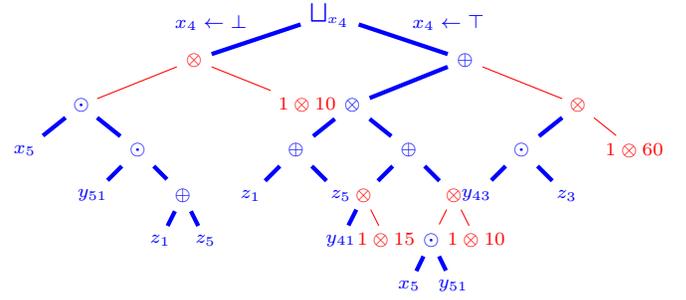

**Figure 6:** D-tree for the semimodule expression $x_4 y_{41}(z_1 + z_5) \otimes 15 +_{\max} x_4 y_{43} z_3 \otimes 60 +_{\max} x_5 y_{51}(z_1 + z_5) \otimes 10$ (part of the annotation of tuple $\langle \text{Gap} \rangle$ in Figure 1e) over the semimodule $\mathbb{B} \otimes \mathbb{N}$. The thick (blue) part is a d-tree for the semiring component of the expression, where $\otimes$ is replaced by $\odot$.

**Decomposition Trees (d-trees)** are a normal form for semiring and semimodule expressions. We next define them and show how to compile arbitrary expressions into d-trees.

DEFINITION 7. *Let $M$ be a monoid and $K$ be a semiring generated by an $S$-valued set $\mathbf{X}$ of variables and constants from $S$. A* decomposition tree, *or d-tree, is a tree, where each inner node is one of $\oplus, \odot, \otimes, \bigsqcup$, or $[\theta]$ and each leaf node is a variable in $\mathbf{X}$ or a constant in $S$, $M$, or $S \otimes M$. The five types of inner nodes have the following meaning.*

*1. A node $\oplus$ with children representing $\Phi$ and $\Psi$ represents the expression $\Phi + \Psi$, where $\Phi$ and $\Psi$ are independent expressions in $K$, or $K \otimes M$ respectively.*

*2. A node $\odot$ with children representing $\Phi$ and $\Psi$ represents the expression $\Phi \cdot \Psi$, where $\Phi$ and $\Psi$ are independent semiring expressions in $K$.*

*3. A node $\otimes$ with children representing $\Phi$ and $\alpha$ represents the expression $\Phi \otimes \alpha$, where $\Phi$ and $\alpha$ are independent expressions in $K$ and $K \otimes M$, respectively.*

*4. A node $[\theta]$ with children representing $\Phi$ and $\Psi$ represents the expression $[\Phi \theta \Psi]$, where $\Phi$ and $\Psi$ are independent expressions in $K$ or $K \otimes M$.*

*5. Given a variable $x \in \mathbf{X}$, an inner node $\bigsqcup_x$ with children representing $\Phi|_{x \leftarrow s_1}, \ldots, \Phi|_{x \leftarrow s_n}$ for all those $s_i \in S$ with $P_x[s_i] \neq 0$ represents the expression $\Phi$.*

Just like semiring or semimodule expressions, d-trees represent probability distributions. Direct implementations of Eq. (4) through (10) are efficient procedures to compute the probability distribution at any inner node of a d-tree, given the probability distributions at its children: Eq. (4) and (6) apply to $\oplus$ nodes, Eq. (5) applies to $\odot$ nodes, Eq. (7) applies to $\otimes$ nodes, Eqs. (8), (9) apply to $[\theta]$, and Eq. (10) applies to $\bigsqcup_x$ nodes. For leaves, we only have the trivial cases of a variable $x$ with probability distribution $P_x$, of constant $s \in S$ or $m \in M$ with distribution $\{(s, 1)\}$ and $\{(m, 1)\}$, and of constant semimodule expressions of the form $s \otimes m$ with probability distribution $\{(s \otimes m, 1)\}$. The probability distribution of the entire d-tree is the distribution of its root and can be computed bottom-up in one pass over the d-tree.

EXAMPLE 12. Consider the d-tree from Figure 5 in which it is assumed that each variable $x \in \{a, b, c\}$ has non-zero probabilities $p_x$ and $\bar{p}_x = 1 - p_x$ for values 1 and 2, respectively. We first compute the probability distribution for the left branch of the d-tree under the aggregation monoid SUM. The distributions are: $\{(1, p_b), (2, \bar{p}_b)\}$ for $b$,

$\{(1,1)\}$ for 1, $\{(2,p_b),(3,\bar{p}_b)\}$ for $b \oplus 1$, $\{(10,1)\}$ for $1 \otimes 10$, $\{(20,p_b),(30,\bar{p}_b)\}$ for $(b \oplus 1) \otimes 10$, $\{(1,p_a),(2,\bar{p}_a)\}$ for $a$, $\{(20,p_ap_b),(30,p_a\bar{p}_b),(40,\bar{p}_ap_b),(60,\bar{p}_a\bar{p}_b)\}$ for $a(b+1)\otimes 10$, $\{(20,1)\}$ for $1\otimes 20$, $\{(40,p_ap_b),(50,p_a\bar{p}_b),(60,\bar{p}_ap_b),(80,\bar{p}_a\bar{p}_b)\}$ for $a(b+1) \otimes 10 + 1 \otimes 20$, and finally

$$\{(40,p_ap_bp_c),(50,p_a\bar{p}_bp_c),(60,\bar{p}_ap_bp_c),(80,\bar{p}_a\bar{p}_bp_c)\}$$

for the entire left branch. For the right branch one obtains

$$\{(70,p_ap_b\bar{p}_c),(80,p_a\bar{p}_b\bar{p}_c),(100,\bar{p}_ap_b\bar{p}_c),(120,\bar{p}_a\bar{p}_b\bar{p}_c)\}.$$

The probability distribution of the entire d-tree is then:

$$\{(40,p_ap_bp_c),(50,p_a\bar{p}_bp_c),(60,\bar{p}_ap_bp_c),(70,p_ap_b\bar{p}_c),$$
$$(80,\bar{p}_a\bar{p}_bp_c + p_a\bar{p}_b\bar{p}_c),(100,\bar{p}_ap_b\bar{p}_c),(120,\bar{p}_a\bar{p}_b\bar{p}_c)\}$$

In case of MIN aggregation, the distribution for both branches as well as for the entire d-tree is $\{(10,1)\}$. In case of the Boolean semiring and MIN aggregation, the distribution for the left branch ($c \leftarrow \bot$) is $\{(10,p_ap_b\bar{p}_c), (\infty,p_a\bar{p}_b\bar{p}_c + \bar{p}_ap_b\bar{p}_c+\bar{p}_a\bar{p}_b\bar{p}_c)\}$, for the right branch ($c \leftarrow \top$) is $\{(10,p_ap_c), (20,\bar{p}_ap_c)\}$, and for the overall d-tree $\{(10, p_ap_b\bar{p}_c + p_ap_c), (20, \bar{p}_ap_c), (\infty,p_a\bar{p}_b\bar{p}_c + \bar{p}_ap_b\bar{p}_c + \bar{p}_a\bar{p}_b\bar{p}_c)\}$. □

We now turn to analysing the complexity of computing the probability distribution for a given d-tree. Since d-trees are binary, the distribution of a convolution node can be computed in time linear in each of the sizes of the distributions of the children according to Eq.(1). The distribution of a $\bigsqcup$-node is computed in time linear in the number of its children and their probability distributions. This implies:

THEOREM 2. *The probability distribution $P_d$ of a d-tree $d$ whose nodes have probability distributions $p_1,\ldots,p_n$ can be computed in time $\mathcal{O}(\prod |p_i|)$.*

For the SUM monoid, the size of $P_d$ can be exponential in the number $n$ of leaves, since there may be exponentially many distinct sums out of $n$ numbers. We analyse two classes of d-trees for which we can obtain a better upper bound on the time complexity of computing $P_d$.

Firstly, observe that the sum of two elements in the MIN or MAX monoid is one of the elements itself (e.g. $a +_{\min} b$ is either $a$ or $b$). This implies that the size of the probability distribution of a semimodule expression $\alpha$ is bounded by the number of monoid values occurring at the leaves of $d$. Moreover, if $\alpha$'s variables are $\mathbb{N}$-valued, $\alpha$'s probability distribution is unaltered under the following *reduction* of its variables to $\mathbb{B}$-valued variables: $P_x[\bot] = P_x[0]$ and $P_x[\top] = 1 - P_x[\bot]$. Hence, one may equivalently consider a d-tree for $\mathbb{B}$-valued variables obtained from this reduction:

PROPOSITION 2. *Given a semimodule expression $\alpha$ over MIN or MAX, the distribution $P_\alpha$ can be computed in time linear in the size of $\alpha$'s d-tree in which all variables are considered by their reduction to $\mathbb{B}$-valued variables.*

Secondly, we analyse SUM aggregation. Evaluating a SUM expression $\sum x_i \otimes m_i$ in which all $x_i$ are independent variables is hard [19]. Yet, instances in which the values $m_i$ are constrained are tractable: we say that a semimodule expression $\sum \phi_i \otimes m_i$ over the SUM monoid $\mathbb{N}$ is *m-bounded* if there is a constant $m \in \mathbb{N}$ such that $\forall i : 0 \leq m_i \leq m$. Aggregation over rationals from a bounded set can be regarded as bounded if the number of decimal places is fixed, e.g. for currencies with 2 decimal places.

```
COMPILE (Expression Φ)
begin
    if Φ has no variables then
     ⌊ return Φ
    if ∃ independent Φ₁, Φ₂ s.t. Φ₁ + Φ₂ = Φ then
     ⌊ return COMPILE(Φ₁) ⊕ COMPILE(Φ₂)
    if ∃ independent Φ₁, Φ₂ s.t. Φ₁ · Φ₂ = Φ then
     ⌊ return COMPILE(Φ₁) ⊙ COMPILE(Φ₂)
    if ∃ independent Φ₁, Φ₂ s.t. Φ₁ ⊗ Φ₂ = Φ then
     ⌊ return COMPILE(Φ₁) ⊗ COMPILE(Φ₂)
    if ∃ independent Φ₁, Φ₂ s.t. [Φ₁θΦ₂] = Φ then
     ⌊ return COMPILE(Φ₁) [θ] COMPILE(Φ₂)
    Choose variable x ∈ X occurring in Φ
    return ⨆ₓ (∀s ∈ S, Pₓ[s] ≠ 0: COMPILE(Φₛ))
```

**Algorithm 1: Compilation of semimodule or semiring expressions into d-trees.**

PROPOSITION 3. *Let $\alpha = \sum_{i=1}^n \phi_i \otimes m_i$ be an m-bounded semimodule expression over the SUM monoid $\mathbb{N}$ where each $\phi_i$ is a product of variables and all variables have non-zero probability only for $0_S$ and $1_S$. The probability distribution $P_d$ of a d-tree $d$ for $\alpha$ can be computed in time $\mathcal{O}(n^2m^2d)$.*

Note that every $\phi_i$ may only evaluate to $0_S$ or $1_S$ and hence the sum is bounded by the product of the number of terms, $n$ and $m$; the product $n \cdot m$ is thus an upper bound for the size of the probability distribution at each node. In particular, COUNT aggregation corresponds to $m_i = 1$ for all $i$. Then the resulting probability distribution has at most size $n$ and can be computed in time $\mathcal{O}(n^2d)$. In a d-tree that contains semimodule expressions $\alpha,\beta$ over different aggregation monoids, the complexity of the sub-trees of $\alpha$ and $\beta$ is each according to Propositions 2 and 3.

**Compiling Expressions into d-trees.**
Algorithm 1 sketches the construction of a d-tree for an input expression $\Phi$ by repeatedly applying six decomposition rules. The obtained d-tree is equivalent to the input expression, i.e., they have the same probability distribution.

The first four rules check whether the input expression can be partitioned and decomposed into two independent expressions $\Phi_1$ and $\Phi_2$. In the first rule, $\Phi_1$ and $\Phi_2$ are either both semimodule or both semiring expressions; in the second rule, they are semiring expressions. In the third rule, $\Phi_1$ is a semiring, and $\Phi_2$ is a semimodule expression.

In order to find independent decompositions in polynomial time, expressions are analysed at a syntactic level: We attempt the first rule only if $\Phi$ is a sum and partition it by the connected components in its clause-dependency graph. In addition to such syntactic manipulations, the decomposition in the second and third rules uses known polynomial-time algorithms to recognise *read-once expressions*, i.e., expressions where each variable occurs once, and hence factorise expressions based on algebraic rewritings such as the associativity and commutativity laws, e.g., [6, 18]. In particular, this approach allows to factor expressions into complex sub-expressions and not only into one variable and the residual.

The last rule decomposes $\Phi$ into sub-expressions $\Phi_s$, for each $s \in S$. As explained for Eq. (10), each expression $\Phi_s$ is obtained in linear time by replacing $x$ with the constant $s$ in $\Phi$. Many heuristics have been proposed for choosing the variable $x$, since good choices can make the difference between polynomial and exponential size decision diagrams (such as ordered binary decision diagrams, d-DNNFs, or d-trees). In our implementation, we choose a variable with most occurrences [18].

REMARK 2. *The requirement that the underlying algebraic structures be commutative and associative is crucial for structural decomposition: without these properties, expression decomposition would be constrained to the fixed order defined by the order of symbols and the nesting of the expression.*

EXAMPLE 13. Figure 5 depicts a d-tree for the semimodule expression $\Phi = a(b+c) \otimes 10 + c \otimes 20$. $\Phi$ cannot be split into independent sub-expressions since variable $c$ occurs in both summands. We choose to eliminate variable $c$ to create mutually exclusive events. (This is a good choice since it leads to independent sub-expressions.) We thus create a node $\bigsqcup_c$ with as many children as valuations of $c$ that have non-zero probability. Consider the case $c \leftarrow 1$ and hence $\Phi|_{c \leftarrow 1} = a(b+1) \otimes 10 + 1 \otimes 20$. This corresponds to the left branch in the d-tree, and can be decomposed using the first rule into its independent summands $a(b+1) \otimes 10$ and $1 \otimes 20$. The former expression can be decomposed using the third rule into independent expressions $a$ and $(b+1) \otimes 10$. The procedure continues until we completely decompose the expressions into variables and semimodule constants.

Figure 6 shows a d-tree for the semimodule expression in the first conditional expression in the annotation $\Phi$ of the result tuple $\langle \text{Gap} \rangle$ in Figure 1e over the semimodule $\mathbb{B} \otimes \mathbb{N}$:

$$x_4 y_{41}(z_1 + z_5) \otimes 15 + x_4 y_{43} z_3 \otimes 60 + x_5 y_{51}(z_1 + z_5) \otimes 10$$

It cannot be partitioned into independent sub-expressions. Choosing variable $x_4$ in the last rule, we create a node $\bigsqcup_{x_4}$. Its left child $\Phi|_{x_4 \leftarrow \bot} = x_5 y_{51}(z_1 + z_5) \otimes 10$ can be decomposed using the third rule into $x_5 y_{51}(z_1 + z_5)$ and $1 \otimes 10$. The former is a read-once expression, on which the second rule can be applied twice to decompose it into $x_5$ and the rest, then the rest into $y_{51}$ and $z_1 + z_5$. The latter is decomposed using the first rule into $z_1$ and $z_5$. The right child

$$\Phi|_{x_4 \leftarrow \top} = y_{41}(z_1 + z_5) \otimes 15 + y_{43} z_3 \otimes 60 + x_5 y_{51}(z_1 + z_5) \otimes 10$$

can be decomposed using a sequence of the first three rules that exploit the independence of sub-expressions.

By construction of query results using $[\![ \cdot ]\!]$, the semiring expression in the second conditional expression in $\Phi$ is precisely the semiring part of the above semimodule expression:

$$x_4 y_{41}(z_1 + z_5) + x_4 y_{43} z_3 + x_5 y_{51}(z_1 + z_5),$$

We can thus use the very same compilation steps as above, with the simplification that only the first two rules, which work on semirings, need be applied. The d-tree is the same as for the semimodule expression, but where $\otimes$ nodes are replaced by $\odot$ and without semimodule expressions at leaves. Figure 6 shows this d-tree with thick (blue) edges. □

PROPOSITION 4. *For any semimodule or semiring expression $\Phi$, Algorithm 1 constructs a d-tree $d$ such that $P_\Phi = P_d$.*

By virtue of this equivalence, we can compute the probability distribution of any expression by first compiling it into a d-tree $d$ and then computing its probability distribution $P_d$. The algorithm is complete since the last rule is always applicable until all variables are evaluated. However, exclusive application of this rule is not desirable: compiling arbitrary expressions by applying the last rule only can lead to d-trees of size exponential in the number of variables. This limitation seems necessary: if we could compile any expression $\Phi$ into a d-tree in polynomial time, then hard problems such as satisfiability, tautology, and even probability computation can be solved in polynomial time for $\Phi$. In practice, however, this rule can be quite effective, as we show in the experiments. In case we need only apply the first four rules, the compilation finishes in polynomial time. This is already known for semiring expressions occurring in the results of tractable relational algebra queries without repeating symbols [18]. In Section 6, we define a fragment of our query language $\mathcal{Q}$ consisting of tractable queries that only create expressions compilable using the first four rules. We finish this section with two observations about our compilation approach, which cannot be presented at length due to space constraints.

**Pruning Conditional Expressions.** The evaluation of $[\alpha \theta \beta]$-expressions can be considerably improved by employing pruning rules that prove parts of $\alpha$ or $\beta$ redundant. Consider the expression $\Phi = [\alpha \theta \beta] = [x \otimes 10 +_{\min} y \otimes 20 \leq 15]$ which can be decomposed into sub-expressions $\alpha$ and $\beta$. The probability $P_\Phi[1_S] = 1 - P_x[0_S]$ is independent of $y$; indeed, when computing $P_\alpha$ one may safely ignore computing $P_\alpha[20]$ since it cannot contribute to $P_\Phi[1_S]$ in the convolution of $[\theta]$. Equivalently, we may replace $\Phi$ with a simpler yet equivalent expression in which redundant terms are pruned. Examples of pruning rules are (similar and symmetric cases omitted):

$$\text{MIN}: \quad \left[ \sum_i \Phi_i \otimes m_i \leq m \right] \equiv \left[ \sum_{i : m_i \leq m} \Phi_i \otimes m_i \leq m \right]$$

$$\text{SUM}: \quad \left[ \sum_i \Phi_i \otimes m_i \leq m \right] \equiv 1_S \quad \text{if } \sum_i m_i \leq m$$

Pruning is particularly effective when the probability distributions of $\alpha$ and $\beta$ have exponential size, such as in case of the SUM monoid; there, early pruning can avoid the full materialisation of such probability distributions.

**Compiling Joint Probability Distributions.** A tuple in the result of an aggregate query in $\mathcal{Q}$ may have several semimodule expressions and be annotated with a conditional expression; in such cases we are interested in finding the joint probability distribution of these expressions. This can be accomplished by compiling the expressions into a single d-tree by applying mutex decomposition until some of the expressions become independent; the joint probability distribution of two independent random variables is simply their product. For instance, given integer variables $a, b, c$ with non-zero probabilities for 1,2 only, the mutex decomposition on $a$ decomposes the joint expression $\langle a+b, a \cdot c \rangle$ into two branches $\langle 1+b, 1 \cdot c \rangle$ and $\langle 2+b, 2 \cdot c \rangle$. In each branch, the two expressions are independent and can be considered separately. For example, the probability for the value $\langle 3, 2 \rangle$ can be worked out to be $P_a[2]P_b[1]P_c[1] + P_a[1]P_b[2]P_c[2]$.

## 6. TRACTABLE QUERY EVALUATION

This section describes the classes $\mathcal{Q}^{\text{ind}}$ and $\mathcal{Q}^{\text{hie}}$ of tractable relational algebra queries with aggregation. The characterisation uses as building blocks queries that return tuple-independent relations and queries reminiscent of hierarchical queries [21]. Class $\mathcal{Q}^{\text{hie}}$ uses $\mathcal{Q}^{\text{ind}}$ as a building block, i.e. $\mathcal{Q}^{\text{ind}} \subset \mathcal{Q}^{\text{hie}}$. Our tractability result follows from the discussion in this section:

THEOREM 3 (TRACTABLE QUERIES).
*Every query in $\mathcal{Q}^{\text{hie}}$ has polynomial-time data complexity.*

The characterisation is based on a generalisation of the hierarchical property for non-repeating conjunctive queries [21]. A query $Q$ is *non-repeating* if every base relation occurs at most once in $Q$; in this section we assume all queries to be non-repeating. Given a query $Q = \pi_{\bar{A}}\sigma_\phi(Q_1 \times \cdots \times Q_n)$ and an attribute $A$, we denote by $A^*$ the set of attributes in $Q$ that are transitively equated with $A$ in $\phi$; $at(A^*)$ denotes the subset of the relation symbols $Q_1, \cdots, Q_n$ in which an attribute from $A^*$ appears. A non-repeating query $Q = \pi_{\bar{A}}\sigma_\phi(Q_1 \times \cdots \times Q_n)$ is *hierarchical* if for each two of its variables $A, B$ that are not in $\bar{A}$ and are not equated with a constant, it holds that $at(A^*) \cap at(B^*) = \emptyset$ or $at(A^*) \supseteq at(B^*)$ or $at(A^*) \subseteq at(B^*)$. Given a query of the above form, an attribute $A$ is a *root attribute* if every relation $Q_1, \ldots, Q_n$ contains some attribute from $A^*$.

Definitions 8 and 9 rely on the following notions and assumptions. Every relation that does not contain semimodule expressions and whose tuples are annotated with distinct variable symbols with given probability distributions is called *tuple-independent*. We allow MIN and MAX aggregation as well as bounded SUM and COUNT aggregation, see Proposition 3. Additionally, we assume that for every aggregation monoid $M$ occurring in a query, the constant $0_M$ does not occur in the database.

We assume that in a selection $\sigma_\phi$, $\phi$ is a conjunction of (1) equality predicates of the form $A=B$ or $A=c$ where $A$ and $B$ are non-aggregation attributes and $c$ is a constant, and (2) $\theta$-comparisons $\alpha\theta\beta$, $\alpha\theta c$, or $\alpha\theta A$ where $\alpha$ and $\beta$ are aggregation attributes and $A$ is a non-aggregation attribute.

We give separate recursive definitions of the classes $\mathcal{Q}^{ind}$ of queries whose result tuples are pairwise independent, and $\mathcal{Q}^{hie}$ whose result tuples may be correlated.

DEFINITION 8 (CLASS $\mathcal{Q}^{ind}$).

1. *Every tuple-independent relation is a $\mathcal{Q}^{ind}$-query.*

2. *Let $Q_1, \ldots, Q_n \in \mathcal{Q}^{ind}$, and $\tilde{Q}_i = \varpi_{\bar{A}_i; \gamma_i \leftarrow AGG_i(C_i)}(Q_i)$. Then the following are $\mathcal{Q}^{ind}$-queries:*

    (a) $Q = \pi_{\bar{A}}\sigma_\phi(\tilde{Q}_1)$, *such that $\gamma_1 \notin \bar{A}$*

    (b) $Q = \pi_{\bar{A}}\sigma_\phi(Q_1 \times \cdots \times Q_n)$, *such that $Q$ is hierarchical and all attributes in $\bar{A}$ are root variables*

    (c) $Q = \pi_\emptyset \sigma_{\gamma_1\theta\gamma_2}(\tilde{Q}_1 \times \tilde{Q}_2)$, *such that $\bar{A}_1 = \bar{A}_2 = \emptyset$.*

Concerning the queries under 8.2, first observe that the tuples in the result of $\tilde{Q}_i$ are pairwise independent and that the expressions $\gamma_i$ have the form $\gamma_i = \sum_{AGG} x_i \otimes v_i$ where the $x_i$ are independent random variables; the annotation of each tuple in $\tilde{Q}_i$ is $\Phi = 1_K$ in case of $\bar{A}_i = \emptyset$ and a sum $\Phi = \sum x_j$ over the annotations of the tuples participating to this group, otherwise. Consider the case 8.2(a): The selection $\sigma_\phi$ may compare $\gamma_1$ with a constant $c$ to yield the annotation $\Phi_t = \Phi \cdot [\gamma_1\theta c]$. Under the assumption that the constant $0_M$ is not in the database, it can be shown that for any valuation $\nu \in \Omega$ it holds that $\nu(\Phi) = 0_S$ if and only if $\nu(\gamma_1) = 0_M$, i.e. the correlation of the distributions of $\Phi$ and $\gamma_1$ is independent of the data. Starting from this observation, it follows that the probability distribution of $\Phi_t = \Phi \cdot [\gamma_1\theta c]$ can by expressed by considering the probabilities of $\Phi$ and $[\gamma_1\theta c]$ separately: For instance, for MIN aggregation and $\theta$ is $\leq$: $P_{\Phi_t}[0_S] = P_{[\gamma_1 \leq c]}[0_S]$ and $P_{\Phi_t}[1_S] = P_{[\gamma \leq c]}[1_S]$. For MIN aggregation and $\theta$ is $\geq$: $P_{\Phi_t}[0_S] = P_\Phi[0_S] + P_{[\gamma_1 \geq c]}[0_S]$ and $P_{\Phi_t}[1_S] = P_\Phi[1_S] + P_{[\gamma \geq c]}[1_S] - 1$. Similar results are obtained for MAX and SUM aggregation. Since the resulting tuples are again independent, the final projection $\pi_{\bar{A}}$ creates as annotations sums of independent expressions; furthermore, as $\gamma_1 \notin \bar{A}$, the only expressions in result tuples are the annotation expressions.

For 8.2(b), it is known that non-repeating hierarchical queries are tractable on tuple-independent relations [21]; since in addition all variables in the projection list are root variables, the result is tuple-independent. In 8.2(c), the conditional expression obtained for the result tuple compares two tractable and independent semimodule expressions.

Secondly, we define $\mathcal{Q}^{hie}$ as follows.

DEFINITION 9 (CLASS $\mathcal{Q}^{hie}$).
*The following are $\mathcal{Q}^{hie}$-queries:*

1. $Q = \pi_{\bar{A}}\varpi_{\bar{A};\gamma \leftarrow AGG(C)}\big[\sigma_\psi(Q_1 \times \cdots \times Q_n)\big]$
   *if $Q_1, \ldots, Q_n \in \mathcal{Q}^{ind}$, and $\pi_{\bar{A}}\sigma_\psi(Q_1 \times \cdots \times Q_n)$ is hierarchical*

2. $Q = \pi_{\bar{A}}\sigma_\phi(Q_1 \times \cdots \times Q_n)$ *if $Q_1, \ldots, Q_n \in \mathcal{Q}^{ind}$, and $Q$ is hierarchical.*

Queries in 9.2 are the well-known non-repeating hierarchical queries [21]. For queries in 9.1, first consider the query $Q' = \pi_{\bar{A}}\sigma_\psi(Q_1 \times \cdots \times Q_n)$ which is by assumption hierarchical. It follows that, given a tuple $t \in Q'$, its annotation $\Phi = \sum \phi_i$ is a read-once expression [17, 21]; moreover, $\Phi$ can be compiled into a d-tree whose size is bounded by the number of its variables. By Propositions 2 and 3, computing the probability distribution of such a d-tree – and thus the probability distribution of $\Phi$ – can be done in time polynomial in the size of $\Phi$. By Proposition 1, the size of the input annotation is polynomial in the size of the input database, hence query evaluation is in polynomial-time.

Now consider the query $Q$ (9.1). Aggregation yields expressions of the form $\alpha = \sum_{AGG} \phi_i \otimes v_i$ where – as above – each $\phi_i$ is a product of variables; since the query is non-repeating, each product has exactly $n$ variables, one from each relation. The aggregated data values $v_i$ are all from the same relation and can hence be associated with the variables from that relation. The expression $\alpha$ can be compiled into the same d-tree as $\Phi$, except that instead of the leaves for the variables $x$ from the aggregated relation, it has leaves of the form $x \otimes v$. The following example illustrates this idea.

EXAMPLE 14. Consider the database in Figure 1 and the query $Q = \varpi_{\emptyset; \alpha \leftarrow SUM(price)}\big(\sigma_{shop='M\&S'}(S) \bowtie PS\big)$ which is of type 1 in $\mathcal{Q}^{hie}$. In this example, the query $Q'$ considered in the explanation above is $Q' = \pi_\emptyset \sigma_{shop='M\&S'}(S) \bowtie PS$; the annotation of its result tuple is $x_1y_{11}+x_1y_{12}+x_2y_{21}+x_2y_{22}+x_3y_{33}+x_3y_{34}$ which is equivalent to the read-once expression $x_1(y_{11} + y_{12}) + x_2(y_{21} + y_{22}) + x_3(y_{33} + y_{34})$. According to the rewriting rules $[\![\cdot]\!]$ in Figure 4, the annotation of the result of $Q$ is $(x_1y_{11}) \otimes 10 + (x_1y_{12}) \otimes 50 + (x_2y_{21}) \otimes 11 + (x_2y_{22}) \otimes 60 + (x_3y_{33}) \otimes 15 + (x_3y_{34}) \otimes 40$. Associating the price values with the annotation variables $y_{ij}$ of their relation, one obtains a read-once expression equivalent to the one above, except that the leaves with $y_{ij}$ variables are now semimodule expressions $y_{ij} \otimes v_i$: $x_1(y_{11} \otimes 10 + y_{12} \otimes 50) + x_2(y_{21} \otimes 11 + y_{22} \otimes 60) + x_3(y_{33} \otimes 15 + y_{34} \otimes 40)$. □

Ré et al. consider queries $\varpi_{\emptyset; \gamma \leftarrow AGG(C)} \sigma_\phi(R_1 \times \cdots \times R_n)$ in which $\pi_\emptyset \sigma_\phi(R_1 \times \cdots \times R_n)$ is hierarchical [19]; these are subsumed by $\mathcal{Q}^{hie}$. For such queries involving aggregation without grouping, the neutral element of the aggregation monoid may safely be in the database without jeopardising query tractability, because the annotation the query result is always $1_K$ and hence the correlation of the distributions of $1_K$ and $\gamma$ is trivial, see also the discussion after Definition 8.

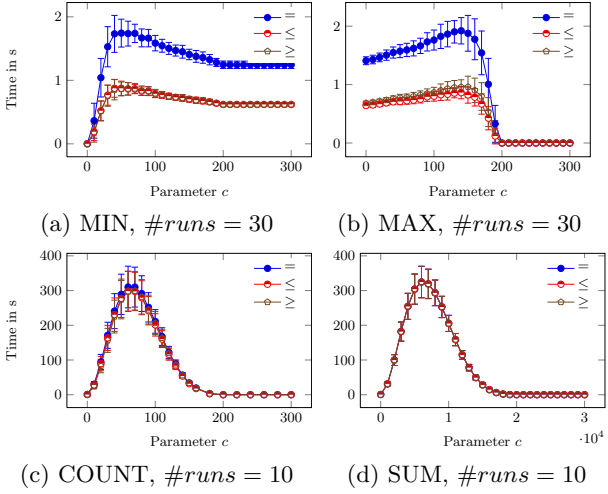

(a) MIN, $\#runs=30$    (b) MAX, $\#runs=30$

(c) COUNT, $\#runs=10$    (d) SUM, $\#runs=10$

**Figure 7: Experiment A: Varying the constant $c$ for different aggregation monoids and comparison operators $\theta$.** $\#v$=25, $L$=200, $R$=0, $\#cl$=3, $\#l$=3, $maxv$=200.

## 7. EXPERIMENTAL EVALUATION

We have assessed our query evaluation technique in two scenarios: randomly generated expressions and aggregate queries on TPC-H data. The results give positive evidence for the feasibility of query evaluation using our technique.

The experiments were performed on a virtual machine with Intel Xeon X5650 Quad 2.67GHz/4GB/64bit running Linux 2.6.35-25/gcc4.4.5. Our algorithms for event construction and probability computation are implemented in C. They are integrated in the SPROUT query engine, which is an extension of PostgreSQL8.3.3. We run each experiment multiple times and report average wall-clock execution times and estimated standard deviation (vertical axis in all figures) while neglecting the slowest and fastest runs.

### 7.1 Experiments on Synthetic Data

We conducted an analysis of the qualitative behaviour of our technique for randomly generated semiring expressions over Boolean random variables of the two forms

$$\left[\sum_{\text{AGGL}}^{L} \Phi_i \otimes v_i \; \theta \; \sum_{\text{AGGR}}^{R} \Psi_j \otimes w_j\right], \; \left[\sum_{\text{AGGL}}^{L} \Phi_i \otimes v_i \; \theta \; c\right] \quad (11)$$

with the following parameters: $L$ ($R$) is the number of semimodule terms on the left (right) of the comparison operator $\theta$, AGGL and AGGR are the aggregation monoids on each side. The second form corresponds to $R$=0. Values $v_i, w_j$ are from $[0, maxv]$ and the expression contains $\#v$ distinct variables. Each $\Phi_i$ has $\#cl$ clauses and each of them has $\#l$ positive literals. In each experiment, we randomly generate $\#runs$ different expressions of form Eq.(11) according to the parameters specified in the respective figures.

**Experiment A** explores the effect of constant $c$ on the evaluation of an expression with $L$=200 terms on the left side compared with $c$ on the right side. Figure 7 depicts the run time for different aggregation and comparison operators. Let us analyse MIN. Values $v_i$ are drawn from $[0, 200]$. For small $c$, our pruning techniques ensure that only terms with values smaller than $c$ are considered by convolution operators in the d-tree. For the operator $\leq$, we thus need to compute the probability that any of these terms is present (i.e., its semiring expression evaluates to true); for the operator $\geq$, we compute the probability that none of these terms is present; for the operator $=$, we compute the probability that none of these terms is present and at least a term with monoid value $c$ is present. The computation becomes slower when increasing $c$, and after $c = 200$, all terms need to be considered, hence the run time converges to a constant value. The behaviour of MAX mirrors that of MIN.

Evaluating COUNT for a constant $c$ effectively amounts to computing the probability that (at most, at least, or exactly) $\binom{L}{c}$ terms are present. The binomial coefficient is trivial for $c = 1$ and $c = L = 200$ and bell-shaped in between. Since we consider three clauses per term ($\#cl = 3$), this experiment evaluates COUNT DISTINCT on top of a conjunctive query. The case of SUM is equivalent to COUNT with the horizontal axis scaled by factor $maxv/2 = 100$. This is to be expected in the limit $\#runs \to \infty$ since the aggregation values are drawn uniformly from $[0, maxv]$ with expected value $maxv/2$. We also considered Experiment A with $L = 100$ terms and the other parameters unaltered. The qualitative behaviour remained unchanged and the run time halved.

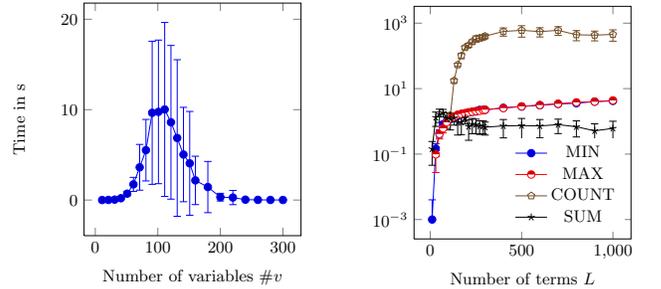

(a) C: $L$=90, $R$=0, $\#cl$=2, $\#l$=2, $maxv$=5, $c$=3, $\theta$ is =, $\#runs$=40, AGGL=MIN.    (b) B: $\#v$=25, $R$=0, $\#cl$=3, $\#l$=3, $maxv$=200, $c$=100, $\#runs$=10, $\theta$ is =.

**Figure 8: Experiments C and B: Varying the number of terms and variables.**

**Experiment B** (Figure 8b) investigates the effect of varying the number of terms while keeping the number of variables constant. The initial increase in run time is due to the cost of partitioning into mutex expressions, which eventually saturates to linear growth for larger expressions once all variables have been considered for partitioning. This experiment mimics answering increasingly complex queries on a database of constant size; this produces increasingly larger expressions, yet with a constant number of variables. We repeated the above experiments (not in the figure) for the remaining comparison operators $\theta \in \{\leq, \geq\}$ and for parameters $maxv = 5, c = 3$ and obtained similar results.

**Experiment C** mirrors Experiment B: Fix the size of the expression and vary the number of its distinct variables. It is known from #SAT analysis that such a setup exhibits an easy/hard/easy phase transition, see Figure 8a. For our algorithm, the transition is understood by its limiting cases: For few variables, expressions can quickly be decomposed into mutex expressions. Conversely, expressions separate into independent sub-expressions in case of many variables since it is likely that different clauses are independent. The large standard deviation in the hard regime suggests that the run time is sensitive to the particular distribution of variables within the expression. In case $\theta$ is $\geq$ or $\leq$, the runtime improves but follows a similar pattern.

**Experiment D** visualises a phase transition similar to that of Experiment C (Figure 9). We explain for Figure 9a:

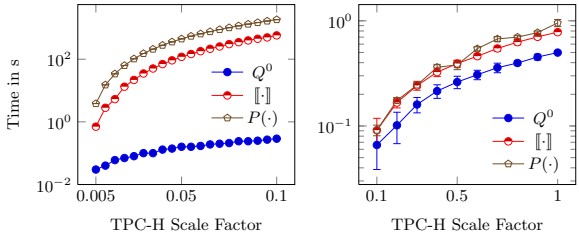

(a) TPC-H Query Q1    (b) TPC-H Query Q2

**Figure 11: Experiment F: Queries on TPC-H Data.**

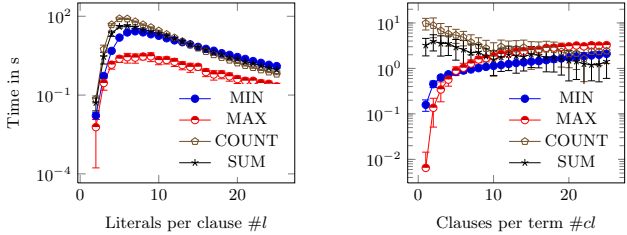

(a) #cl=3    (b) #l=3

**Figure 9: Experiment D: Varying the number of literals per clause (a) and of clauses per term (b).** #v=25, L=100, R=0, maxv=5, c=3, #runs=20, $\theta$ is $\leq$.

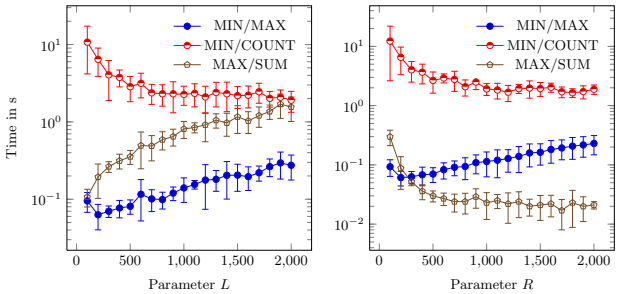

(a) R=150, L variable    (b) L=150, R variable

**Figure 10: Experiment E: Expressions with different left/right aggregations.** #v=25, #cl=2, #l=2, maxv=200, c=100, #runs=10, $\theta$ is $\leq$.

When keeping the number of variables and terms constant while increasing the arity of clauses, the problem becomes easy for small and large clauses, and hard in between.

**Experiment E** investigates the behaviour of expressions with different aggregation operators on each side (Figure 10). We analyse the case $\sum_{\text{MAX}} \leq \sum_{\text{SUM}}$. When increasing the number of terms on the left/MAX side (left figure), it becomes more likely that the maximum value on the left side is larger than the sum on the right side; hence more terms have to be compiled. When increasing the number of terms on the right/SUM side (right figure), already a few mutex decomposition steps satisfy enough clauses to make the sum larger than the maximum on the left side, i.e. the compilation is faster with increasing number of SUM terms.

## 7.2 Queries on TPC-H

We consider tuple-independent TPC-H 2.14.0 databases of scales up to 1GB, and two TPC-H queries. The query $Q_1$ reports the amount of business that was billed, shipped, and returned (only the COUNT aggregate is selected). The query $Q_2$ is a join of five relations and with a nested aggregate query, which asks for suppliers with minimum cost for an order for a given part in a given region. For each query, we compared the execution times (1) on a deterministic database ($Q^0$) without expression or probability computation, (2) of the computation of the expressions ($[\![\cdot]\!]$), and (3) of probability computation for the result tuples ($P(\cdot)$).

**Experiment F** compares the runtime for $Q^0$, $[\cdot]$, and $P(\cdot)$ (Figure 11). The overhead of expression computation and probability computation is polynomial, because the TPC-H dataset scales up the amount of tuples while keeping tuple correlations (i.e. the number of tuples directly related via joins within a group) constant. The performance difference between $Q_1$ and $Q_2$ is due to the selectivity of the queries: The size of the annotation expressions as a measure for the number of tuples contributing to $Q_1$'s answer is 1200 times larger than the size of $Q_2$'s annotations.

**Acknowledgments.** This research was funded by the ERC FP7 grant agreement FOX number FP7-ICT-233599 and EPSRC grant PrOQAW.

## 8. REFERENCES

[1] S. Abiteboul, T.-H. H. Chan, E. Kharlamov, W. Nutt, and P. Senellart. Aggregate Queries for Discrete and Continuous Probabilistic XML. In *ICDT*, 2010.
[2] Y. Amsterdamer, D. Deutch, and V. Tannen. Provenance for Aggregate Queries. In *PODS*, 2011.
[3] P. Buneman and W. Tan. Provenance in Databases. In *SIGMOD*, 2007.
[4] D. Burdick, P. M. Deshpande, T. S. Jayram, R. Ramakrishnan, and S. Vaithyanathan. OLAP over Uncertain and Imprecise Data. *VLDBJ*, 16(1), 2006.
[5] A. Darwiche and P. Marquis. A Knowledge Compilation Map. *JAIR*, 2002.
[6] M. C. Golumbic, A. Mintz, and U. Rotics. An Improvement on the Complexity of Factoring Read-once Boolean Functions. *Discrete Applied Mathematics*, 156(10), 2008.
[7] T. J. Green, G. Karvounarakis, and V. Tannen. Provenance Semirings. In *PODS*, 2007.
[8] G. Grimmett and D. Welsh. *Probability - An Introduction*. Oxford University Press, 1986.
[9] L. Han. Evaluation of Aggregate Queries in Probabilistic Databases. MSc Thesis, University of Oxford, 2011.
[10] R. Jampani, F. Xu, M. Wu, L. L. Perez, C. Jermaine, and P. J. Haas. MCDB : A Monte Carlo Approach to Managing Uncertain Data. In *SIGMOD*, 2008.
[11] B. Kanagal, J. Li, and A. Deshpande. Sensitivity Analysis and Explanations for Robust Query Evaluation in Probabilistic Databases. In *SIGMOD*, 2011.
[12] O. Kennedy and C. Koch. PIP: A Database System for Great and Small Expectations. In *ICDE*, 2010.
[13] C. Koch. Incremental Query Evaluation in a Ring of Databases. In *PODS*, 2010.
[14] C. Koch and D. Olteanu. Conditioning Probabilistic Databases. *PVLDB*, 1(1), 2008.
[15] J. Lechtenbörger, H. Shu, and G. Vossen. Aggregate Queries over Conditional Tables. *JIIS*, 19(3), 2002.
[16] R. Murthy, R. Ikeda, and J. Widom. Making Aggregation Work in Uncertain and Probabilistic Databases. *TKDE*, 2011.
[17] D. Olteanu and J. Huang. Using OBDDs for Efficient Query Evaluation on Probabilistic Databases. In *SUM*, 2008.
[18] D. Olteanu, J. Huang, and C. Koch. Approximate Confidence Computation in Probabilistic Databases. In *ICDE*, 2010.
[19] C. Ré and D. Suciu. The Trichotomy of HAVING Queries on a Probabilistic Database. *VLDBJ*, 18(5), 2009.
[20] M. Soliman, I. Ilyas, and K. C.-C. Chang. Probabilistic top-k and Ranking-Aggregate Queries. *TODS*, 33(3), 2008.
[21] D. Suciu, D. Olteanu, C. Ré, and C. Koch. *Probabilistic Databases*. Morgan & Claypool Publishers, 2011.
[22] M. Yang, H. Wang, H. Chen, and J. Ku. Querying Uncertain Data with Aggregate Constraints. In *SIGMOD*, 2011.